\documentclass[apj,twocolumn,twocolappendix,numberedappendix]{openjournal}
\usepackage{newtxtext,newtxmath,enumitem,multirow}
\usepackage[utf8]{inputenc}
\usepackage[T1]{fontenc}
\usepackage[breaklinks,colorlinks,allcolors=blue,citecolor=blue,urlcolor=blue]{hyperref}
\usepackage{orcidlink}
\usepackage{ctable}
\pdfoutput=1

\def\Msun{\, M_{\odot}}

\shorttitle{Effect of the LMC on the kinematics of satellites}
\shortauthors{Kravtsov \& Winney}

\begin{document}
\title[Effect of the LMC on the kinematics of satellites]{Effect of the Large Magellanic Cloud on the kinematics of Milky Way satellites\\ and virial mass estimate\vspace{-1.5cm}}

\author{Andrey Kravtsov\,\orcidlink{0000-0003-4307-634X}$^{1,2,3,\star}$}
\author{Sophia Winney\,\orcidlink{0009-0004-9634-4808}$^{1}$\vspace{1mm}}
\affiliation{$^{1}$Department of Astronomy  \& Astrophysics, The University of Chicago, Chicago, IL 60637 USA}
\affiliation{$^{2}$Kavli Institute for Cosmological Physics, The University of Chicago, Chicago, IL 60637 USA}
\affiliation{$^{3}$Enrico Fermi Institute, The University of Chicago, Chicago, IL 60637 USA}
\thanks{$^\star$\href{mailto:kravtsov@uchicago.edu}{kravtsov@uchicago.edu}}


\begin{abstract}
We present a study illustrating the effects of the passage of a Large Magellanic Cloud (LMC) mass satellite on the distance and velocity distributions of satellites in $\Lambda+$Cold Dark Matter simulations of Milky Way (MW) sized halos. In agreement with previous studies, we find that during such a passage the velocity distribution develops a high-velocity tail, which can bias velocity-based virial halo mass estimates. When the velocity distribution of MW satellites is corrected for effects of the LMC passage, it is consistent with the distributions in halos of masses as low as $M_{\rm 200c}=8\times 10^{11}\, \Msun$ and as high as $1.5\times 10^{12}\,\Msun$. We present a new halo mass estimator $M_{\rm 200c}=c\sigma^2_{\rm 3D}r_{\rm med}$, where $c$ is the coefficient calibrated using satellite systems in the simulated MW-sized halos, $\sigma^2_{\rm 3D}$ is the variance of 3D velocities taken with the sign of the radial velocity of each satellite, and $r_{\rm med}$ is the median halocentric distance of the satellites. We show that the estimator has only $s=8\%$ scatter around the median relation of the estimated and true halo masses and deviates by $<2s$ from the median during the pericentric passage of an LMC-like subhalo. This is because $\sigma^2_{\rm 3D}$ and $r_{\rm med}$  deviate in the opposite directions during such passages. We apply the estimator to the MW satellite system and estimate the virial mass of the Milky Way of $M_{\rm 200c}=9.96\pm 1.45\times 10^{11}\, \Msun$, in good agreement with several recent estimates using other methods.\vspace{2mm}

\end{abstract}


\maketitle



\section{Introduction}
\label{sec:intro}

Velocities of galaxies have been long used to infer information about the mass of virialized objects, such as galaxy groups and clusters \citep{Zwicky.1933,Zwicky.1937,Smith.1936}, unvirialized outskirts of such objects \citep[e.g.,][]{Diaferio.etal.2005} and gravitationally interacting unvirialized pairs of galaxies \citep{Kahn.Woltjer.1959,Li.White.2008,Gonzalez.etal.2014,Penarrubia.etal.2016}.

Velocities of galaxies moving in and around the Milky Way (MW) galaxy have also been used to probe its mass under the assumption that such galaxies have negligible tangential velocity or are bound \citep{Sandage.1986,Zaritsky.etal.1989,Kochanek.1996}.  \citet[][cf. also \citealt{Sohn.etal.2013}]{Boylan.Kolchin.etal.2013}, for example, used the velocity of motion of the Leo I galaxy in the outskirts of the MW halo to constrain the virial mass of the Milky Way to $>10^{12}\, M_\odot$ at the 95\% confidence level. 
At the same time, modeling of the Sagittarius dwarf stream indicated a Milky Way halo mass of $<10^{12}\, M_\odot$ \citep{Gibbons.etal.2014}.

Such differences can be expected because the use of instantaneous velocity and position of individual satellite galaxies to estimate host halo mass is prone to biases if the satellite is in an atypical part of its orbit \citep{Patel.etal.2017a,Patel.etal.2017b}.
Using the velocities of an ensemble of satellite galaxies can mitigate some of the biases. \citet{Watkins.etal.2010} present an estimator of halo mass and use radial velocities of the MW satellites to estimate MW halo mass within 300 kpc distance, but find that the mass estimate is very sensitive to the assumption about velocity anisotropy of the satellite system. At the time, only a limited number of tangential velocity estimates existed and thus estimates of the MW halo mass focused on the Bayesian methods incorporating constraints from individual satellites using the phase-space  distribution function calibrated with cosmological simulations of MW-sized halos \citep{Patel.etal.2017a,Patel.etal.2017b,Patel.etal.2018,Eadie.etal.2017}. 

The dramatic increase in the number of tangential velocity measurements for many MW dwarf satellites and globular clusters with the advent of the Gaia satellite data \citep{McConnachie.Venn.2020,McConnachie.Venn.2020b,Li.etal.2021,Battaglia.etal.2022} allowed statistical analyses of the velocity distributions using much larger ensembles of objects \citep{Callingham.etal.2019,Eadie.Juric.2019,Vasiliev.2019,Posti.Helmi.2019,Fritz.etal.2020,Li.etal.2020,Rodriguez_Wimberly_etal2022}.
Similarly, accurate measurements of phase-space information for stars in the Sagittarius dwarf spheroidal stream and measurements of the LMC velocity allowed for the identification of a clear misalignment between the stream track and the direction of the reflex-corrected proper motions in the leading arm of the stream. \citet{Vasiliev.etal.2021} showed that this misalignment is due to the interaction between the LMC and MW, and its modeling allowed them to accurately estimate the total halo masses of both objects.

The improved velocity accuracy and sophistication of the analyses led to reduced uncertainties in halo mass estimates and reduced discrepancies between different studies. Nevertheless, recent MW halo virial mass measurements still allow a fairly wide range of values $\approx 6-20\times 10^{11}\, M_\odot$  \citep[e.g.,][]{Deason.etal.2021,Vasiliev.etal.2021,Correa_Magnus.Vasiliev.2022,Koposov.etal.2023,Zhou.etal.2023,Roche.etal.2024}.  

One systematic uncertainty in the inferences of halo mass from the positions and velocities of dwarf satellite galaxies and globular clusters is the effect of the Large Magellanic Cloud (LMC) on the orbits and velocities of such tracers \citep[e.g.,][see \citealt{Vasiliev.2023} for a review]{Vasiliev.etal.2021,Pace.etal.2022}. Effects of the LMC on the nearby stellar streams and dwarf galaxy orbits indicate its halo mass of $1.3\pm 0.3\times 10^{11}\, M_\odot$ \citep{Erkal.etal.2019,Vasiliev.etal.2021,Shipp.etal.2021,Correa_Magnus.Vasiliev.2022,Koposov.etal.2023,Watkins.etal.2024,Sheng.etal.2024} or $\approx 15\%$ of the MW halo mass.

Such a large LMC mass results in perturbations of velocities and orbits of individual MW satellites \citep{Erkal.etal.2020,Pace.etal.2022,Makarov.etal.2023}, significant motions of the inner Milky Way, and asymmetries of the potential
in which dwarf satellites and orbit \citep{Gomez.etal.2015,Erkal.etal.2020,Garavito.Camargo.etal.2019,Garavito-Camargo.etal.2021,DSouza.Bell.2022}. In particular, LMC induces the uniform reflex motion of the Milky Way disk relative to the outer halo and distorts the distribution of velocities. The latter leads to negative mean radial velocity of the tracers, such as halo stars, globular clusters, and satellites. 

These effects have been detected using different tracers in the Milky Way halo \citep{Petersen.Penarrubia.2020,Petersen.Penarrubia.2021,Erkal.etal.2021,Conroy.etal.2021,Yaaqib.etal.2024}.  
In particular, LMC both induces the motion of the MW disk relative to the outer halo and affects the velocities of the outer satellites, such as Antlia II \citep{Ji.etal.2021} and Leo I \citep{Erkal.etal.2020}. The use of the latter in particular tended to produce higher halo mass estimates in prior analyses. \citet{Erkal.etal.2020} also showed that the mass estimator of \citet{Watkins.etal.2010} applied to the velocities of satellites and stellar tracers overestimates the enclosed mass $M(<r)$ by up to $50\%$ at large $r$ if the LMC-induced effects are neglected. Likewise, \citet{Deason.etal.2021} showed that the motion induced by LMC bias mass estimates based on the 6D phase space information of halo stars. The presence of the bound Magellanic Clouds pair also affects statistical estimates of the Milky Way halo mass \citep{Gonzalez.etal.2013}.
Finally, the approach of the LMC has a large effect on the spatial distribution of MW satellites \citep{Nadler.etal.2020,Manwadkar.Kravtsov.2022,Garavito.Camargo.etal.2023}. 

Large effects of the LMC on the phase-space distribution of the potential Milky Way tracers need to be accounted for in theoretical inferences of halo mass that use these tracers. This is challenging, however, because only $\approx 5-10\%$ of the MW-sized halos have a pair of LMC-like satellites with a pericenter close to $\approx 50$ kpc \citep{Boylan.Kolchin.etal.2010,Busha.etal.2011}. Thus, large samples of halos need to be simulated with a high resolution sufficient to resolve and model satellite evolution reliably. Alternatively, the analyses can focus on the properties of tracers that are not significantly affected by the dynamical perturbations due to the LMC, such as specific angular momentum \citep{Patel.etal.2018}.

In this study, we explore the effects of the LMC on the velocity distribution of satellite systems using a Caterpillar suite of high-resolution simulations of MW-sized halos \citep[][see Section \ref{sec:modelling}]{Griffen.etal.2016}. We demonstrate that similarly to the real LMC, LMC-like objects in simulations induce motions onto their host halo and its satellites which creates a high-velocity tail in the distribution of halocentric 3D satellite velocities (Section \ref{sec:vdist_lmc}). We also show how accounting for this tail affects the conclusion about MW halo mass based on the velocity distribution. We then present a new halo mass estimator that has only $\approx 8\%$ scatter around the true mass when applied to simulated halos (Section \ref{sec:m200est}) and which is affected by the LMC-like objects by only $\lesssim 10\%$. We use the estimator calibrated with the simulated halos to infer the MW halo mass of $M_{\rm 200c}=9.93\pm 1.45\times 10^{11}\,M_\odot$.  We summarize our results and conclusions in Section~\ref{sec:summary}.

\section{Modeling Milky Way satellite system}
\label{sec:modelling}

We model the population of Milky Way dwarf satellite galaxies around the Milky Way using tracks of haloes and subhaloes from the Caterpillar \citep{Griffen.etal.2016} suite of $N$-body simulations\footnote{\url{https://www.caterpillarproject.org}} of 32 MW-sized haloes.  We use the highest resolution suite LX14 to maximize the dynamic range of halo masses probed by our modelling. 
 
The haloes were identified using the modified version of the Rockstar halo finder and the Consistent Trees Code \citep{Behroozi.etal.2013}, with modification improving recovery of the subhaloes with high fraction of unbound particles \citet[see discussion in Section 2.5 of][]{Griffen.etal.2016}. As was shown in \citet[][see their Fig. 1]{Manwadkar.Kravtsov.2022}, the subhalo peak mass function in the LX14 simulations is complete at $\mu=M_{\rm peak}/M_{\rm host} \gtrsim 4 \times 10^{-6}$, or $M_{\rm peak} \approx 4 \times 10^{6} \Msun$ for the host halo mass $M_{\rm host}\approx 10^{12}\, M_\odot$, even in the innermost regions of the host ($r < 50$ kpc).  This is sufficient to model the full range of luminosities of observed Milky Way satellites, as faintest ultrafaint dwarfs are hosted in haloes of $M_{\rm peak}\gtrsim 10^7\, M_\odot$ in our model \citep[][]{Kravtsov.Manwadkar.2022,Manwadkar.Kravtsov.2022}. 

The mass evolution tracks of subhaloes of MW-sized host haloes are used as input for the \texttt{GRUMPY} galaxy formation model, which evolves various properties of gas and stars of the galaxies they host. It is a regulator-type galaxy formation framework \citep[e.g.,][]{Krumholz.Dekel.2012,Lilly.etal.2013,Feldmann.2013} designed to model galaxies of $\lesssim L_\star$ luminosity \citep[][]{Kravtsov.Manwadkar.2022} which follows the evolution of several key galaxy properties by solving a system of coupled different equations. 

The model accounts for UV heating after reionization and associated gas accretion suppression onto small mass haloes, galactic outflows, a model for gaseous disk and its size, molecular hydrogen mass, star formation, etc.  The evolution of the half-mass radius of the stellar distribution is also modeled. The galaxy model parameters used in this study are identical to those used in \citet{Manwadkar.Kravtsov.2022}. 
Importantly for this study, the model was shown to reproduce the luminosity function and radial distribution of the Milky Way satellites and the size-luminosity relation of observed dwarf galaxies \citep{Manwadkar.Kravtsov.2022}. In \citet{Pham.etal.2023} we also showed that the flattening of the subhalo distribution in the Caterpillar suite is consistent with observed flattened distribution of Milky Way satellites. 
Here we use the model to predict luminosities and stellar masses of satellite galaxies around the MW-sized haloes from the Caterpillar suite. 

\begin{figure}
  \centering
  \includegraphics[width=0.49\textwidth]{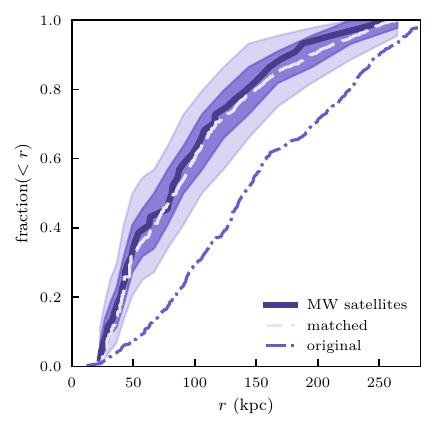}
  \caption[]{The cumulative distribution (CDF) of distances of satellites to the center of their parent halo for the Milky Way satellites (thick solid line). The dark and light-shaded regions show the $68.2$ and $95.6$ percentiles of the CDFs for the bootstrap resamples of the MW satellite distances at a given distance $d$. The dot-dashed line shows the cumulative distance distribution of subhalos of a peak mass of $M_{\rm peak}>10^8\, M_\odot$ in the main host halos in the Caterpillar simulations. The light long-dashed line shows the cumulative distribution of the distances for subhalos selected using the distance matching procedure described in the text (Section \ref{sec:vdist_lmc}). The figure shows that the matching procedure selects subhalos with a distance distribution statistically consistent with that of the Milky Way satellites.}
   \label{fig:cdfd}
\end{figure}

\begin{figure}
  \centering
  \includegraphics[width=0.49\textwidth]{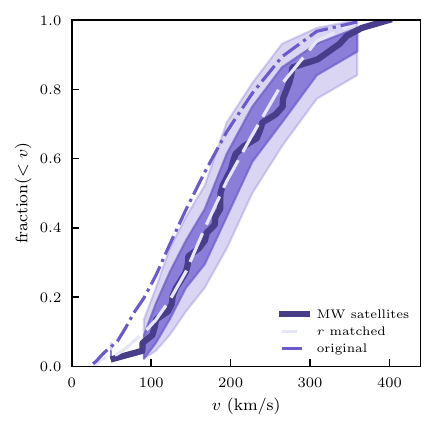}
  \caption[]{The cumulative distribution of 3D velocities of satellites in the Milky Way and simulated halos. The velocity distribution of the Milky Way satellites is shown by the thick solid line. The dark and light-shaded regions show the $68.2$ and $95.6$ percentiles of the CDFs for the bootstrap resamples of the MW satellite velocities at a given velocity $v$. The dot-dashed line shows the cumulative distance distribution of velocities for subhalos of the peak mass of $M_{\rm peak}>10^8\, M_\odot$ in the main host halos in the Caterpillar simulations. The light long-dashed line shows the cumulative distribution of the velocities for subhalos selected using the distance matching procedure described in the text (Section \ref{sec:vdist_lmc}). The figure shows that the matching procedure selects subhalos with a velocity distribution statistically consistent with that of the Milky Way satellites.}
   \label{fig:cdfv_dmatch}
\end{figure}

\begin{figure}
  \centering
  \includegraphics[width=0.49\textwidth]{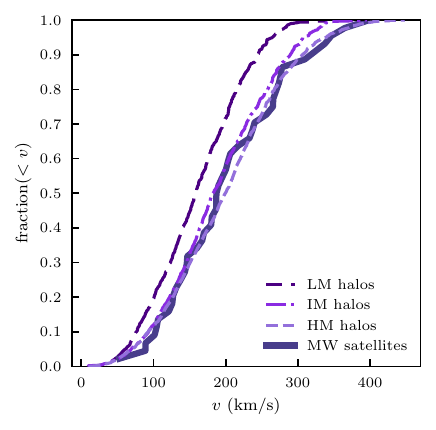}
  \caption[]{The cumulative distribution of 3D velocities of satellites in the Milky Way and simulated halos of different halo mass ranges. The velocity distribution of the Milky Way satellites is shown by the thick solid line. The long-dashed, dot-dashed, and dashed lines show velocity CDFs of satellites in the low, intermediate, and high mass host halo mass ranges. The satellites have been selected using the distrance distribution matching procedure described in the text (Section \ref{sec:vdist_lmc}). }
   \label{fig:cdfv_mbins}
\end{figure}

\begin{figure}
  \centering
  \includegraphics[width=0.49\textwidth]{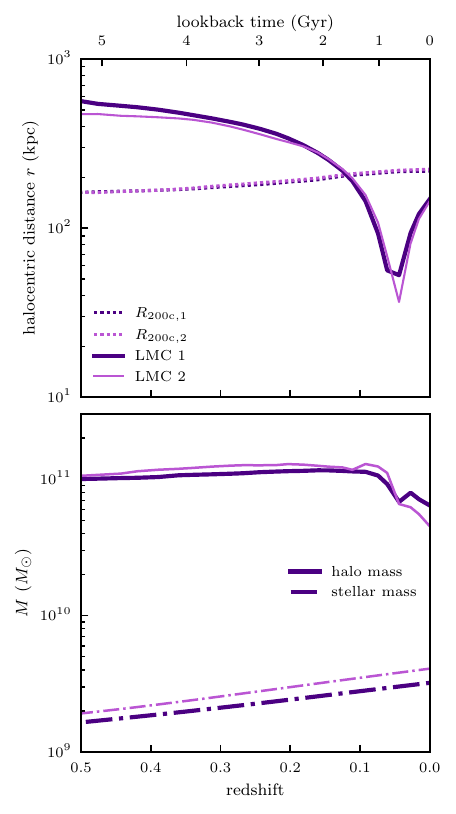}
  \caption[]{{\it Top panel:}\/ evolution of distance (in physical kpc) of two LMC-like satellites to the center of their host halo from $z=0.5$ to $z=0$ in two different Caterpillar host halos. The satellites are on their first infall and reach their pericenter of $\approx 40-50$ kpc at $z_{\rm LMC,peri}\approx 0.04-0.05$. The dotted lines show evolution of the virial radii $R_{\rm 200c}(z)$ of their host halos. The upper $x$-axis shows the lookback time corresponding to the redshifts shown in the bottom $x$-axis. {\it Bottom panel:}\/ Evolution of halo masses (solid lines) and stellar masses (dashed lines) of the two LMC-like satellites shown in the top panel. The satellites reach peak halo masses of $M_{\rm peak}\approx 1.16\times 10^{11}\,M_\odot$ and $M_{\rm peak}\approx 1.35\times 10^{11}\,M_\odot$ at $z\approx 0.1$ before they start to lose mass due to tidal stripping. The stellar mass is computed using our galaxy formation model and continues to grow due to ongoing star formation and reaches $M_\star\approx 3\times 10^9\, M_\odot$, comparable to the stellar mass of the observed LMC.}
   \label{fig:dm_z}
\end{figure}

\begin{figure}
  \centering
  \vspace{1cm}
  \includegraphics[width=0.5\textwidth]{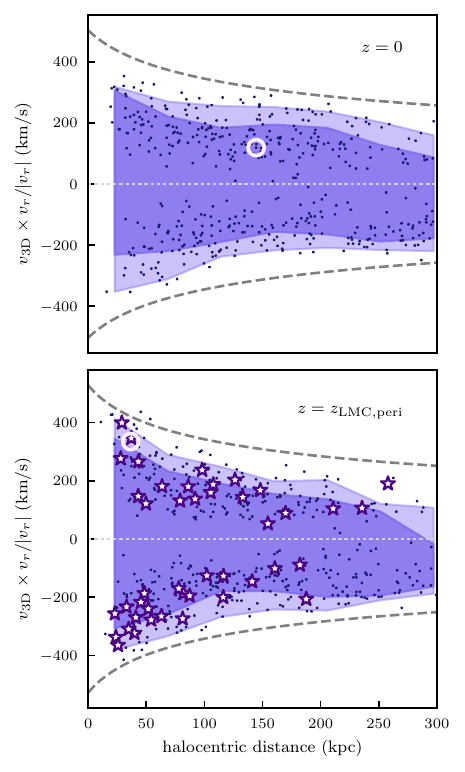}
  \caption[]{Distribution of satellite subhalos (dots) in the plane of halocentric distance and 3D velocity in one of the host halos in which an LMC-like pericentric passage occurs at $z_{\rm LMC,peri}\approx 0.05$. The velocities are taken with the sign of the radial velocity of each satellite. The dark and light-shaded bands show the $15.9-84.1\%$ and $2.3-97.7\%$ ranges of subhalo velocity distribution in distance bins. The large circle in each panel shows the location of the LMC-like satellite. The gray dashed lines show escape velocity profiles for the NFW halo with the same $M_{\rm 200c}$ and concentration as the simulated halo at each redshift. 
  The {\it lower panel\/} shows the distribution at $z=z_{\rm LMC,peri}$ and includes the Milky Way dwarf galaxy satellites (stars) for comparison. For reference, the real LMC is at $r\approx 50$ kpc and a total velocity of $320$ km/s \citep[e.g.,][]{Patel.etal.2018} and has just passed its pericenter.
  The {\it upper panel\/} shows the velocity distribution at $z=0$, which is approximately symmetric around zero, while at $z_{\rm LMC,peri}$ it significantly shifted towards negative velocities.}
   \label{fig:vd_lmc_effect}
\end{figure}
%
\section{Results}
\label{sec:results}

\subsection{Effect of the LMC on the velocity distribution of satellites}
\label{sec:vdist_lmc}

For comparison of Caterpillar-simulated subhalos to Milky Way satellites, we use the distance and velocity measurements found in Table 1 of \citet[][hereafter RW22]{Rodriguez_Wimberly_etal2022}. Galactocentric tangential velocities and heliocentric radial velocities are taken from \citet{McConnachie.2020a} and \cite{McConnachie.2020b}, while the heliocentric distance and error measurements are from \citet{Simon.2019}, \citet{Karachentsev.Kashibadze.2006}, \citet{Weisz.2016}, and \citet{Torrealba.2019}. The measurements are converted to the Galactocentric reference frame in RW22.

Comparison to the Milky Way requires a satellite population whose cumulative Galactocentric distance distribution matches that of the MW satellites. To obtain this, we draw a set of random distances from the MW satellite distance distribution using the inverse transform sampling method and spline approximation of the inverse cumulative distribution function of MW satellite distances. We then selected subhalos in each simulated MW-sized halo that were closest to each of the drawn distances. Figure \ref{fig:cdfd} shows that this distance-matching technique produces a satellite population with cumulative distance distribution consistent with the distance distribution of MW the satellites. Note that both here and below the distance $r$ is the distance to the center of the host halo defined by the halo finder (corresponding to the central density peak). At the same time, Figure \ref{fig:cdfv_dmatch} demonstrates that the 3D velocity distribution of the distance-matched subhalos is statistically consistent with the velocity distribution of Milky Way satellites. 

We also checked the effect of distance-matching on the cumulative distributions of satellite velocity's radial and tangential components. Distance matching generally slightly worsens the agreement of the model and observed radial velocity distributions, but the effect of the matching on the radial velocity distribution is small. Distance matching does improve agreement of the tangential velocity distributions and it is this improvement that is responsible for the better match of the 3d velocity distribution. 

Following RW22, we divided the simulated host halos into three mass bins: low mass ($<10^{12}\ \Msun$), intermediate mass ($1-1.2 \times 10^{12}\ \Msun$), and high mass ($>1.2 \times 10^{12}\ \Msun$). Figure \ref{fig:cdfv_mbins} shows the cumulative velocity distribution of satellites in halos in each mass bin and compares them to the velocity distribution of the Milky Way satellites. The intermediate and high mass bins are closest to the distribution of the Milky Way, while the low mass bin is shifted towards lower velocities. Our results are somewhat different from the results shown in Figure 3 of RW22 where the distribution of velocities in halos in the high mass bin is the least consistent with the Milky Way. This is likely due to the difference in the mass distribution of halos in the high-mass bin. The high-mass halos in our sample contain masses in the range $1.2-1.6\times 10^{12}\, M_\odot$, while the sample of RW22 lacks halos in the $1.2-1.4 \times 10^{12}\ \Msun$ range but contains halos of mass up to $1.9 \times 10^{12}\ \Msun$. The velocity distribution of satellites in their high-mass bin is thus shifted to higher velocities compared to our sample.

Hosts 4 and 53 from the Caterpillar host halo tracks include the passage of subhalos that reach peak masses of $M_{\rm peak} = 1.16 \times 10^{11}\, \Msun$ and $M_{\rm peak} = 1.35 \times 10^{11}\, \Msun$, respectively. These are comparable to the mass of the LMC with $M_{\rm peak} = 1.3\pm 0.3\times 10^{11}\, M_\odot$ \citep{Erkal.etal.2019,Vasiliev.etal.2021,Shipp.etal.2021,Watkins.etal.2024}. The time evolution of the halocentric distance and mass of these subhalos are shown in Figure~\ref{fig:dm_z}. The satellites reach their peak masses $M_{\rm peak}\approx 1.16\times 10^{11}\,M_\odot$ and $M_{\rm peak}\approx 1.35\times 10^{11}\,M_\odot$ at $z \approx 0.1$ and begin to lose mass due to tidal stripping during their pericentric passage at $z \approx 0.05$ or about $\approx 0.7$ Gyr after they cross $R_{\rm 200c}$ of their host halo. This is similar to the estimates of the accretion time of the LMC in the Milky Way \citep{Petersen.Penarrubia.2020}. As the LMC-like satellites continue to be tidally stripped, stellar mass steadily increases due to continued star formation and reaches a peak of $M_\star\approx 3\times 10^9\, M_\odot$, comparable to the observational estimates of the LMC's stellar mass.

\begin{figure}
  \centering
  \includegraphics[width=0.49\textwidth]{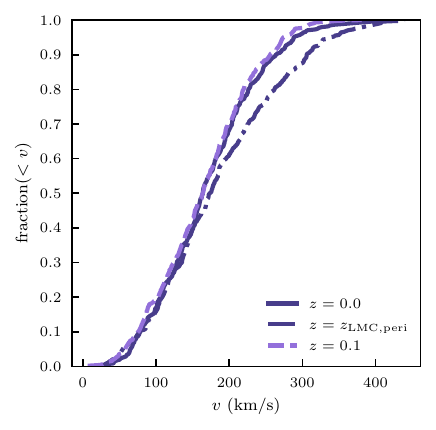}
  \includegraphics[width=0.49\textwidth]{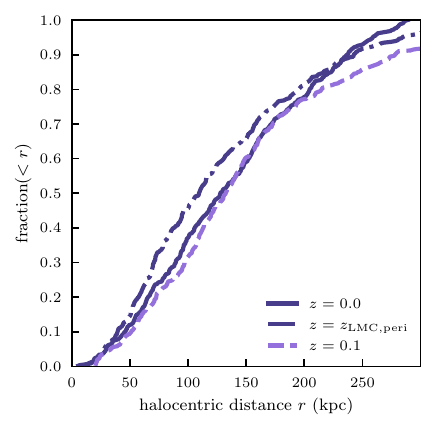}
  \caption[]{Effect of the LMC on the cumulative distribution of 3D velocities $v=\vert v_{\rm 3d}\vert$ (upper panel) and halocentric distances (lower panel) of satellite subhalos. Dashed lines show the CDFs of satellites of the main host at $z=0.1$ when the LMC-like satellite was still $\gtrsim 150$ kpc away from the center, while solid lines show the CDFs of satellites at $z=0$ when the LMC-like satellite is again at $d\approx 150$ kpc (see Figure \ref{fig:dm_z} where this LMC-like object is labeled as LMC1). The dot-dashed lines show the CDFs at $z_{\rm LMC,peri}=0.05$ when the LMC-like satellite was close to the pericenter. The velocity distributions are for satellites not selected to match the distance distribution of the MW satellites. Note that the distribution at $z_{\rm LMC,peri}$ has a significant tail of high-velocity satellites, which is not present in the velocity CDFs at $z=0.1$ and $z=0$. The LMC thus significantly skews the velocity distribution towards higher velocities as it passes near the pericenter of its orbit. At the same time, the distribution of distances is skewed towards smaller distances at the LMC pericentric passage.}
   \label{fig:cdfvd_lmc_effect}
\end{figure}

\begin{figure*}
  \centering
  \includegraphics[width=0.49\textwidth]{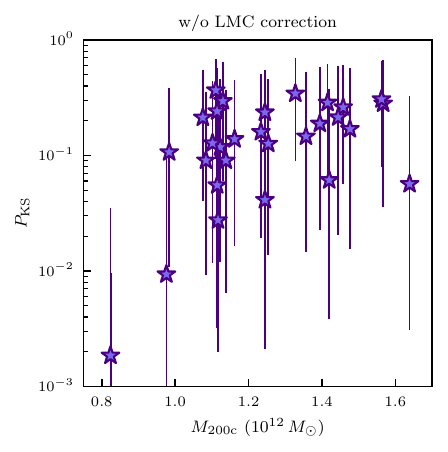}
  \includegraphics[width=0.49\textwidth]{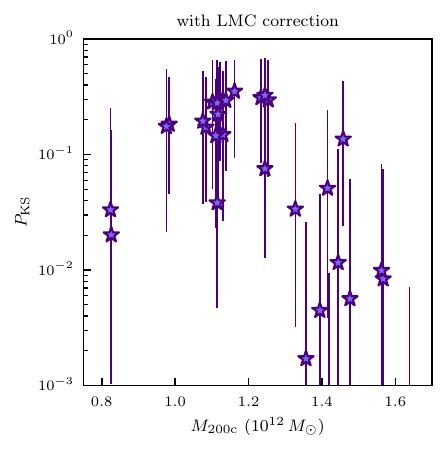}
  \caption[]{The distributions of the Kolmogorov-Smirnov test $p$-values for MW-sized halos in the sample. The vertical lines show the 68 percentile of the distribution of $p$-values obtained for bootstrap resamples of the satellites in the simulated halos and the Milky Way satellites. The star symbols show the medians of the distributions.   {\it Left panel} shows the $p$-values computed using velocity distributions not corrected for the effect of the LMC, while {\it right panel} shows the case when simulated satellite velocities have been corrected to account for the effect of the LMC. The correction significantly diminishes the $p$-values for halos of $M_{\rm 200c}>1.3\times 10^{12}\, M_\odot$, but they are still sufficiently high for most halos to be statistically consistent with the Milky Way. In all cases, satellites in simulated halos are selected to match the distance distribution of the MW satellites.}
   \label{fig:pks_lmc}
\end{figure*}

\begin{figure*}
  \centering
  \includegraphics[width=0.49\textwidth]{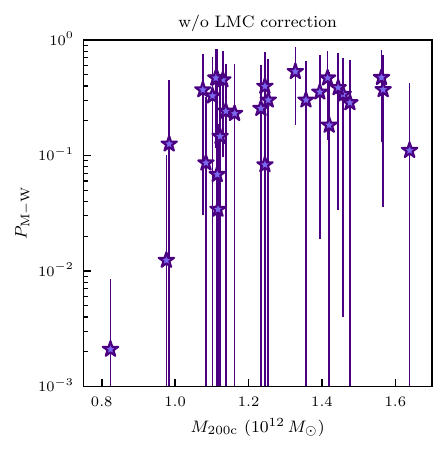}
  \includegraphics[width=0.49\textwidth]{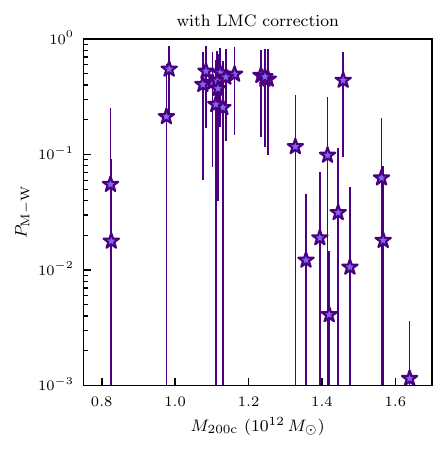}
  \caption[]{The distributions of the Mann-Whitney $U$ test $p$-values for MW-sized halos in the sample. The vertical lines show the 68 percentile of the distribution of $p$-values obtained for bootstrap resamples of the satellites in the simulated halos and the Milky Way satellites. The star symbols show the medians of the distributions.  {\it Left panel} shows the case without taking into account the effect of the LMC on the velocity distribution of satellites, while {\it right panel} shows the case when simulated satellite velocities have been corrected to account for the effect of the LMC. The correction significantly diminishes the $p$-values for halos of $M_{\rm 200c}>1.3\times 10^{12}\, M_\odot$. As for the KS test $p$-values in Figure \ref{fig:pks_lmc}, the correction significantly diminishes the $p$-values for halos of $M_{\rm 200c}>1.3\times 10^{12}\, M_\odot$, but they are still sufficiently high for most halos to be statistically consistent with the Milky Way. In all cases, satellites in simulated halos are selected to match the distance distribution of the MW satellites.}
   \label{fig:pmw_lmc}
\end{figure*}

\begin{figure}
  \centering
  \includegraphics[width=0.49\textwidth]{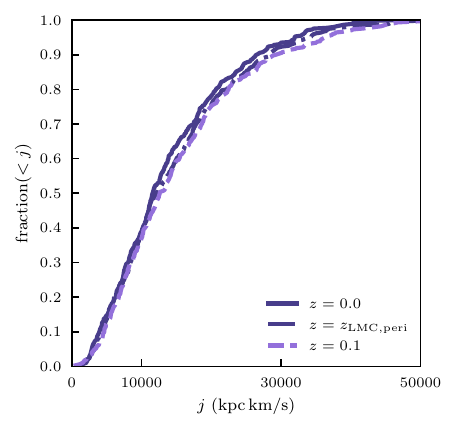}
  \includegraphics[width=0.49\textwidth]{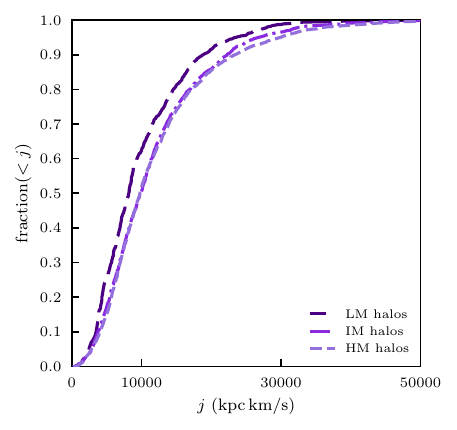}
  \caption[]{Upper panel: effect of the LMC on the cumulative distribution function of specific angular momentum of satellites $j=\vert \mathbf{v}\times \mathbf{r}\vert$. The distributions are shown at $z=0.1$ (dashed line), when the LMC-like satellite was still $\gtrsim 150$ kpc away from the center, and at $z=0$, when the LMC-like satellite is again at $d\approx 150$ kpc (solid line). The dot-dashed line shows the CDF at $z_{\rm LMC,peri}=0.05$ when the LMC-like satellite was close to the pericenter. Unlike velocity distribution, the specific angular momentum distribution changes little during pericentric passage of the LMC-sized halo. {\it Lower panel:} specific angular momentum distribution of satellites in the MW-sized host halos in the low- ($M_{\rm 200c}\leq 10^{12}\,M_\odot$), intermediate ($M_{\rm 200c}\in (1-1.2]\times 10^{12}\,M_\odot$), and high-mass bins ($M_{\rm 200c}>1.2 10^{12}\,M_\odot$). The dependence of the $j$ distribution on halo mass is weaker than the dependence of the velocity distribution.}
   \label{fig:cdfj_lmc_effect}
\end{figure}

For one of the host halos, Figure \ref{fig:vd_lmc_effect} shows the satellite velocity distribution at $z = 0$ and at the time of pericentric passage $z \approx 0.05$, as well as the 15.9 - 84.1 and 2.3 - 97.7 percentile intervals of the velocity distribution in bins of distance. The effect of the LMC-like satellite is greatest near the pericenter, where the subhalo velocity distribution becomes skewed toward larger negative velocities. At $z = 0$, the velocities return to a symmetric distribution. The effect of the LMC-like subhalo passage on the cumulative velocity distribution is shown in Figure \ref{fig:cdfvd_lmc_effect}, which shows that the satellite distribution has a tail of higher velocities near the pericentric passage. The cumulative velocity distribution at $z = 0$, on the other hand, is similar to the distribution before the LMC-like subhalo entered the host halo at $z = 0.1$. 

The lower panel of Figure \ref{fig:vd_lmc_effect} shows that the cumulative distance distribution of satellites becomes more centrally concentrated during the pericentric passage of the LMC-like subhalo. Although the physical mechanism of the distance distribution change warrants a detailed investigation, it is most likely a result of contraction of the satellite orbits in response to the deepening of the potential due to the LMC mass when it is near the pericenter. Although the LMC pericenter is $\approx 50$ kpc, the virial radius of the LMC and its mass is about 100 kpc and encompasses and affects a significant fraction of the Milky Way's virial volume.  Like the velocity distribution, the distance distribution at $z = 0$ becomes closer to the distribution before the passage at $z = 0.1$, although the distribution at $r\gtrsim 180$ kpc remains somewhat different most likely due to the presence of the LMC counterpart at $r=150$ kpc at $z=0$ (see Figure~\ref{fig:dm_z}). 

The physical origin of the large change in the velocity and distance distribution of satellites during the pericentric passage of an LMC-like subhalo is the reflex motion of the central region of the MW-sized host halo and associated compression of satellite distribution and velocities in response to the gravitational pull of the massive subhalo and due to a deepening of the potential when the LMC counterpart reaches the pericenter of its orbit. 

We have computed both the overall mean 3D velocity of reflex motion of the simulated satellites relative to the halo center \citep[$\mathbf{v}_{\rm travel}$ in][]{Petersen.Penarrubia.2020,Petersen.Penarrubia.2021} and the mean radial velocity of satellites after the $x$, $y$, and $z$ components of the mean 3D velocity are subtracted from the velocity components of the satellites. The reflex motion velocity ($v_{\rm travel}$) for the two host halos with LMC-like satellites is 44 and 38 km/s. The mean radial velocities after the mean velocity subtraction are $\langle v_r\rangle = -24$ km/s and $-44$ km/s. These numbers are comparable to the corresponding mean 3D velocity of motion of the inner Milky Way and mean radial velocity of satellites estimated by \citet[][see their Table 1]{Petersen.Penarrubia.2021}. Note that the mean negative radial velocity of satellites reflects a distortion of their velocity distribution by the LMC in addition to the overall mean reflex motion. Such distortion (dubbed ``compression'' by the authors) is now reliably detected in the velocity tracers of the Milky Way \citep{Petersen.Penarrubia.2021,Yaaqib.etal.2024}. We find that in simulations by $z=0$ when LMC counterparts move to $r\approx 150$ kpc, both the reflex motion velocity and the mean radial velocity decrease by a factor of three.

The large effect of the LMC on the velocity and distance distribution needs to be taken into account in the estimates of MW halo mass \citep{Erkal.etal.2020}. Although the effect becomes small well after the passage, the LMC in the Milky Way has just passed its pericenter and therefore effects on the MW satellite velocities are expected to be significant. To illustrate the effect of change in velocity distribution on halo mass inference, we approximate the difference in the velocity distributions at the pericentric passage of the LMC-like subhalo and at $z=0$ as a function of 3D velocity of simulated satellites using a cubic spline. We then use the spline to compute the correction for each satellite of the Milky Way using its 3D velocity.  Note that this correction is estimated using only two MW-sized host halos with LMC-like satellites that are present in the Caterpillar sample and is thus uncertain. We use it here for illustration of the potential effects of the LMC on the virial mass inference. 

Figure \ref{fig:pks_lmc} shows the Kolmogorov-Smirnov (KS) $p$-values for the consistency between the velocity distributions of distance-matched satellite samples in simulated host halos of different masses and the Milky Way satellites with and without applying this correction. The figure shows the medians of the $p$-values obtained for the bootstrap resamples of the velocities of the distance-match subhalo and MW satellite samples, while vertical lines show the extent of the 68th percentiles of the $p$-value distribution of these samples. The figure shows that if the effect of the LMC is neglected, low halo masses are disfavored by the KS test, while intermediate and high-mass halos are consistent with the MW distribution. When the velocity correction is applied, the $p$-values of the high-mass halos decrease markedly and the distribution of $p$-values instead peaks in the intermediate halo masses. 

The KS test is sensitive to the differences of both the mean shift and the overall shapes of the distribution. 
RW22 used the Mann-Whitney $U$ test, which is primarily sensitive to differences in the median of distributions. 
Figure \ref{fig:pmw_lmc} shows the results of the Mann-Whitney test with and without correction for the LMC effect.  The points and error bars show medians and 68th percentiles of the $p$-values estimated for the bootstrap resamples of the velocities of MW satellites and distance-matched satellite subhalos in the simulated host halos.  The figure shows results similar to the KS test results in Figure \ref{fig:pks_lmc}. 
Without the correction, the $p$-values are high for halos with $M_{\rm 200c}\gtrsim 10^{12}\, M_\odot$, while smaller mass halos are disfavored. When the LMC correction is applied to velocities, the $p$-values for halos with $M_{\rm 200c}\gtrsim 1.3\times 10^{12}\, M_\odot$ decrease substantially, although the overall range of $p$-values still does not exclude these virial masses.

We note that when the distribution of the $p$-values for bootstrap resamples is considered, the velocity distribution does not exclude any halos in the mass range of the Caterpillar suite regardless of whether the LMC correction to velocities is applied. The correction only lowers the $p$-values of halos with $M_{\rm 200c}\gtrsim 1.3\times 10^{12}\, M_\odot$. Thus, the velocity distribution is not a sensitive indicator of halo mass. 

\citet[][see also \citealt{Patel.etal.2017b}]{Patel.etal.2018} used the specific angular momenta of satellites to estimate MW halo mass and showed that they are not sensitive to deviations from dynamic equilibrium. The upper panel of Figure \ref{fig:cdfj_lmc_effect} shows the cumulative distribution of specific angular momentum $j=\vert \mathbf{r}\times \mathbf{v}_{\rm 3D}\vert$ at three different epochs, $z=0.1$, $z=z_{\rm LMC,peri}$ and $z=0$, for one of the host halos with an LMC-like satellite passing through its pericenter at $z_{\rm LMC,peri}\approx 0.05$. The effect of the LMC-like satellite on the $j$ distribution is indeed very small, which is consistent with the conclusion of \citet{Patel.etal.2018}. Thus, the use of the angular momentum to estimate halo mass would not require a correction for the effect of the LMC. 

At the same time, the lower panel of Figure \ref{fig:cdfj_lmc_effect} shows that the cumulative $j$ distributions for halos in the three mass bins are not sufficiently different to derive an accurate mass estimate. 
\citet{Patel.etal.2018} find that the median specific angular momentum, $j_{\rm med}$ of satellites in halos scales as  $j_{\rm med}\propto M_{\rm 200c}^{0.78\pm 0.03}$. Given the halo mass range of the Caterpillar host halo sample, we would expect $j_{\rm med}$ to differ at least by a factor of $\approx 1.5$ between the LM and HM mass bins. However, the median in the lower panel of Figure \ref{fig:cdfj_lmc_effect} varies less than this and there is little difference in the distributions of $j$ for the IM and HM samples. We think this is most likely due to the wide span of $j$ distribution for hosts of a given mass. When dealing with a finite number of host halos and their satellite, specific subsamples from this wide distribution can result in a large scatter of $j_{\rm med}$ that obscures the average trend of $j_{\rm med}$ with $ M_{\rm 200c}$. 

These considerations indicate that both velocity and specific angular momentum distributions quantified using small samples of simulated halos do not allow an accurate estimate of MW halo mass.  This motivated us to search for an alternative proxy of $M_{\rm 200c}$ which is both sensitive to the value of the mass and robust against effects of the LMC.

\begin{figure}
  \centering
  \includegraphics[width=0.49\textwidth]{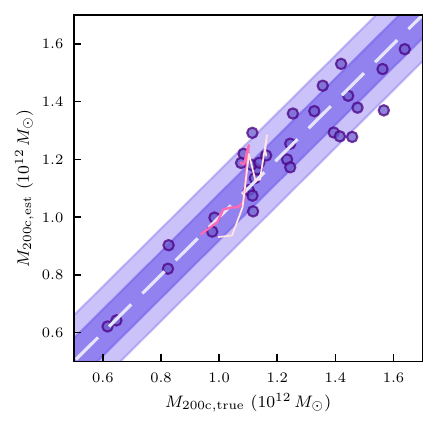}
  \caption[]{True halo mass $M_{\rm 200c}$ for the 32 host halos in the Caterpillar sample vs the mass $M_{\rm 200c,est}$ estimated using the median satellite distance within 300 kpc, the velocity dispersion of their 3D velocity distribution, and the constant determining using the true $M_{\rm 200c}$ of the Caterpillar MW-sized host halos. The white long-dashed line is a one-to-one relation, while dark and light-shaded bands show the $1\sigma$ ($\approx 0.08$) and $2\sigma$ scatter of the sample estimated around this relation. The two thin lines show the evolution of the halo mass and its estimate from $z=0.15$ to $z=0$ for the two host halos in which the LMC-like satellite passes through its pericenter at $z_{\rm LMC,peri}\approx 0.04-0.05$. Masses estimated for the MW-sized halos using eq. \ref{eq:m200est} in the sample exhibit a tight linear relation with their true masses with only $\approx 8\%$ scatter. }
   \label{fig:m200true_m200est}
\end{figure}

\begin{figure}
  \centering
  \includegraphics[width=0.49\textwidth]{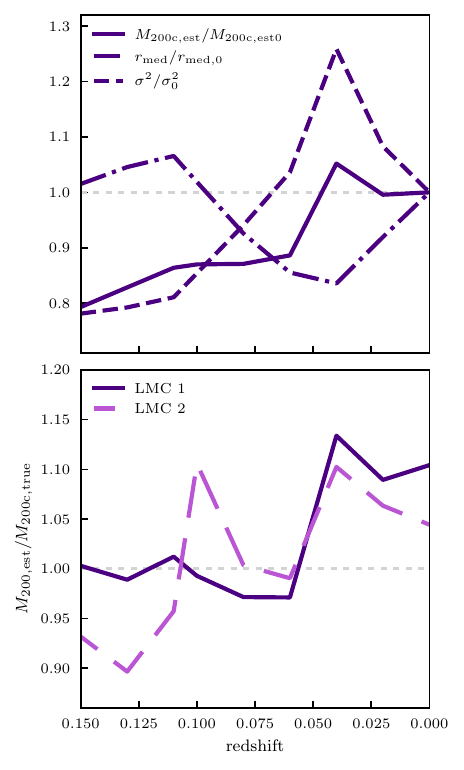}
  \caption[]{{\it Upper panel:} effect of the LMC on the median halocentric distance of satellites within 300 kpc of the host (dot-dashed line), variance of the 3D velocity $\sigma^2$, and the halo mass estimate from $z=0.15$ to $z=0$ in the halo, in which pericentric passage of the LMC-sized object on the first infall occurs at $z\approx 0.04-0.05$ when distance to the host is $\approx 50$ kpc (see Fig. \ref{fig:dm_z}). All quantities are normalized to their values at $z=0$. Note that the LMC induces an upward fluctuation of $\sigma^2$ but a downward fluctuation of $d_{\rm med}$. These fluctuations thus offset each other in the estimate $M_{\rm 200c,est}$ using eq. \ref{eq:m200est}, which exhibits a fluctuation of only $\approx 10\%$. {\it Lower panel}: the ratio of the halo mass estimated using equation \ref{eq:m200est} to its true mass during the LMC-sized object's pericentric passage. Prior to the pericentric passage, eq. \ref{eq:m200est} provided an unbiased estimate of $M_{\rm 200c}$. During and immediately after the passage, the estimate is biased by $\approx 5-10\%$. }
   \label{fig:dmsig_lmc_effect}
\end{figure}

\subsection{Robust halo mass estimator }
\label{sec:m200est}

The mass of a system in equilibrium should be proportional to a measure of the size of the system and the velocity variance of some tracers of the halo potential. We therefore searched for an estimator of this kind, taking into account that during the LMC passage, the distance distribution of the satellites shifts to smaller distances, while velocity distribution shifts to larger velocities. We find that an accurate estimator of the halo mass enclosing the density contrast of 200 relative to the critical mass, $M_{\rm 200c}$, is given by\footnote{Alternatively, one can recast this expression as $M_{\rm 200 c, est}=c^\prime \,G^{-1}r_{\rm med}\,\sigma^2_{\rm 3D}$, where $c^\prime=1.226\pm 0.012$ and the gravitational constant $G$, $r_{\rm med}$ and $\sigma^2_{\rm 3D}$ are in cgs units.}
\begin{equation}
M_{\rm 200 c, est}=c \,r_{\rm med}\,\sigma^2_{\rm 3D},
\label{eq:m200est}
\end{equation}
where the coefficient of proportionality $c=2.85\pm 0.028 \times 10^5\,\Msun$ was estimated using the true values of $M_{\rm 200c}$ of the 32 Caterpillar host halos, $r_{\rm med}$ is the median halocentric distance of the satellites within $r_{\rm max}=300$ kpc in kiloparsecs, and $\sigma^2_{\rm 3D}$ is the variance of $v_{\rm 3D}\times \mathbf{v}_r/\vert v_r\vert$ for satellites (i.e. velocities plotted in Fig.~\ref{fig:vd_lmc_effect}) within $r_{\rm max}$ in kilometers per second. Note that the variance of velocities with the sign of radial velocities is used here, not just the variance of the absolute values of $v_{\rm 3D}$. This variance measures the width of the 3D velocity distribution of halo satellites shown in Figure \ref{fig:vd_lmc_effect}. 

Figure \ref{fig:m200true_m200est} demonstrates that the accuracy of this halo mass estimator is not significantly impacted by the passage of the LMC. The ratio of the halo mass obtained using the estimator to the halo mass given by Caterpillar for each host shows only $s\approx 8\%$ scatter around the 1:1 ratio. The light lines show the evolution of the estimated versus true mass for the two host halos with an LMC-like satellite from before its infall at $z = 0.15$ to $z = 0$, well after its pericentric passage. The estimator stays within the $2s$ range through the duration of the satellites' passages, indicating that the estimator is not overly sensitive to the effects of an LMC-sized satellite. 

The upper panel of Figure \ref{fig:dmsig_lmc_effect} shows the evolution of the factors of eq. \ref{eq:m200est} compared to their values at $z = 0$ as a function of redshift. At $z = 0.04-0.05$, when the LMC-sized satellite is closest to the host, the median halocentric distance of the satellites decreases while the variance of the 3D velocity distribution increases. This reflects the changes in the velocity distribution and distance cumulative distribution shown in Figures \ref{fig:vd_lmc_effect} and \ref{fig:cdfvd_lmc_effect}. As a result, the value of the estimated mass deviates from its $z = 0$ estimate by less than 10\% during the pericentric passage of the LMC-like satellite at $z\approx 0.04$. 

The lower panel of Figure \ref{fig:dmsig_lmc_effect} shows the evolution of the ratio of the estimated halo mass to the true mass from $z=0.15$ to $z=0$. It shows that during and after the pericentric passage, the estimated mass is $\approx 10\%$ higher than the true mass. Thus, given that the scatter around the relation is $\approx 8\%$, the LMC passage biases the mass by $1.25\sigma$. This means that the mass estimator of eq. \ref{eq:m200est} is robust against effects of the LMC passage. The small 10\% correction due to the effect of the LMC can be applied to the mass estimate of the Milky Way.

With the proportionality constant obtained from the simulated halos, using eq. \ref{eq:m200est} with all of the Milky Way satellites with measured 3D velocities and applying 10\% correction for the effect of the LMC yields a Milky Way halo mass of $M_{\rm 200c}=9.96\pm 1.45\times 10^{11}\,M_\odot$, where uncertainty is estimated using bootstrap resampling. If we use only the classical satellites of the Milky Way ($M_V<-8$) we get a consistent estimate $M_{\rm 200c}=8\pm 2.4\times 10^{11}\,\Msun$, albeit with a larger uncertainty. 

\section{Summary and conclusions}
\label{sec:summary}

We presented a study illustrating the effects of the passage of LMC-mass satellites on the distance and velocity distributions of satellites in the Milky Way-sized halos. We also considered the effect of such passages on their mass estimates. Our results and conclusions are as follows. 

\begin{itemize}
    \item[1.] In agreement with results of \citet{Erkal.etal.2020} and \citet{Deason.etal.2021}, we find that the effect of the passage of an LMC-like satellite on the velocity distribution is substantial and this effect can bias mass estimates based on velocities. We show that during the pericentric passage, the velocity distribution develops a high-$v$ tail due to the motions induced by LMC, while the satellite spatial distribution becomes more centrally concentrated. 
    \item[2.] We show that correcting for the effect of the LMC on the 3D velocity distribution of satellites when using this distribution to estimate halo mass favors halo masses in the range $1-1.3\times 10^{12}\,\Msun$. However, we also show that the velocity distribution is a rather low accuracy estimator of halo mass as velocity distributions in halos of masses as low as $M_{\rm 200c}=8\times 10^{11}\, \Msun$ and as high as $1.5\times 10^{12}\,\Msun$ are still statistically consistent with the velocity distribution of the MW satellites. 
    \item[3.] We considered the distribution of specific angular momenta $j$ as an estimator of mass and confirmed that the $j$ distribution is not sensitive to the passage of LMC-mass satellites. At the same time, we find that the $j$ distribution is a weak function of host halo mass and is therefore a relatively poor mass indicator.
    
    \item[4.] We present a new halo mass estimate that is proportional to the variance of 3D velocities taken with the sign of the radial velocity of each satellite and the median halocentric distance of the satellites (eq. \ref{eq:m200est}). The velocity variance is a measure of the width of the velocity distribution shown in Figure \ref{fig:vd_lmc_effect}. The estimator has only $8\%$ scatter around the mean $M_{\rm 200c,est}-M_{\rm 200c}$ relation.
    
    \item[5.] We show that the velocity variance and median distance deviate in opposite directions during the pericentric passage of the LMC-like satellites (Fig. \ref{fig:dmsig_lmc_effect}). For this reason, the estimator  deviates by $<2s$, where $s$ is the scatter around the median $M_{\rm 200c,est}-M_{\rm 200c}$ relation during the pericentric passage of the LMC-like subhalos.
    
    \item[6.] We apply the estimator to the MW satellite system and estimate the virial mass of the Milky Way to be $M_{\rm 200c}=9.96\pm 1.45\times 10^{11}\, \Msun$, in good agreement with several recent estimates using other methods \citep[e.g.,][]{Erkal.etal.2020,Deason.etal.2021,Vasiliev.etal.2021,Correa_Magnus.Vasiliev.2022}. 

\end{itemize}

\section*{Acknowledgements}
We are grateful to Vasily Belokurov, Denis Erkal and the UChicago structure formation and Cambridge Streams group for useful discussions during this project.
We thank Alexander Ji and the Caterpillar collaboration for providing halo tracks of the Caterpillar simulations used in this study.
AK was supported by the National Science Foundation grant AST-1911111 and NASA ATP grant 80NSSC20K0512. SW was supported by the University of Chicago CCRF’s Quad Research Scholarship program.
Analyses presented in this paper were greatly aided by the following free software packages: {\tt NumPy} \citep{numpy_ndarray}, {\tt SciPy} \citep{scipy}, {\tt Matplotlib} \citep{matplotlib}, and \href{https://github.com/}{\tt GitHub}. We have also used the Astrophysics Data Service (\href{http://adsabs.harvard.edu/abstract_service.html}{\tt ADS}) and \href{https://arxiv.org}{\tt arXiv} preprint repository extensively during this project and the writing of the paper.

\section*{Data Availability}

Halo catalogs from the Caterpillar simulations are available at \href{https://www.caterpillarproject.org/}{\tt https://www.caterpillarproject.org/}. The \texttt{GRUMPY} model pipeline is available at \url{https://github.com/kibokov/GRUMPY}. The data used in the plots within this article are available on request to the corresponding author.

\bibliographystyle{mnras}
\bibliography{lmc}

\begin{thebibliography}{}
\makeatletter
\relax
\def\mn@urlcharsother{\let\do\@makeother \do\$\do\&\do\#\do\^\do\_\do\%\do\~}
\def\mn@doi{\begingroup\mn@urlcharsother \@ifnextchar [ {\mn@doi@}
  {\mn@doi@[]}}
\def\mn@doi@[#1]#2{\def\@tempa{#1}\ifx\@tempa\@empty \href
  {http://dx.doi.org/#2} {doi:#2}\else \href {http://dx.doi.org/#2} {#1}\fi
  \endgroup}
\def\mn@eprint#1#2{\mn@eprint@#1:#2::\@nil}
\def\mn@eprint@arXiv#1{\href {http://arxiv.org/abs/#1} {{\tt arXiv:#1}}}
\def\mn@eprint@dblp#1{\href {http://dblp.uni-trier.de/rec/bibtex/#1.xml}
  {dblp:#1}}
\def\mn@eprint@#1:#2:#3:#4\@nil{\def\@tempa {#1}\def\@tempb {#2}\def\@tempc
  {#3}\ifx \@tempc \@empty \let \@tempc \@tempb \let \@tempb \@tempa \fi \ifx
  \@tempb \@empty \def\@tempb {arXiv}\fi \@ifundefined
  {mn@eprint@\@tempb}{\@tempb:\@tempc}{\expandafter \expandafter \csname
  mn@eprint@\@tempb\endcsname \expandafter{\@tempc}}}

\bibitem[\protect\citeauthoryear{{Battaglia}, {Taibi}, {Thomas}  \&
  {Fritz}}{{Battaglia} et~al.}{2022}]{Battaglia.etal.2022}
{Battaglia} G.,  {Taibi} S.,  {Thomas} G.~F.,   {Fritz} T.~K.,  2022, \mn@doi
  [\aap] {10.1051/0004-6361/202141528}, \href
  {https://ui.adsabs.harvard.edu/abs/2022A&A...657A..54B} {657, A54}

\bibitem[\protect\citeauthoryear{{Behroozi}, {Wechsler}, {Wu}, {Busha},
  {Klypin}  \& {Primack}}{{Behroozi} et~al.}{2013}]{Behroozi.etal.2013}
{Behroozi} P.~S.,  {Wechsler} R.~H.,  {Wu} H.-Y.,  {Busha} M.~T.,  {Klypin}
  A.~A.,   {Primack} J.~R.,  2013, \mn@doi [\apj] {10.1088/0004-637X/763/1/18},
  \href {https://ui.adsabs.harvard.edu/abs/2013ApJ...763...18B} {763, 18}

\bibitem[\protect\citeauthoryear{{Boylan-Kolchin}, {Springel}, {White}  \&
  {Jenkins}}{{Boylan-Kolchin} et~al.}{2010}]{Boylan.Kolchin.etal.2010}
{Boylan-Kolchin} M.,  {Springel} V.,  {White} S. D.~M.,   {Jenkins} A.,  2010,
  \mn@doi [\mnras] {10.1111/j.1365-2966.2010.16774.x}, \href
  {https://ui.adsabs.harvard.edu/abs/2010MNRAS.406..896B} {406, 896}

\bibitem[\protect\citeauthoryear{{Boylan-Kolchin}, {Bullock}, {Sohn}, {Besla}
  \& {van der Marel}}{{Boylan-Kolchin} et~al.}{2013}]{Boylan.Kolchin.etal.2013}
{Boylan-Kolchin} M.,  {Bullock} J.~S.,  {Sohn} S.~T.,  {Besla} G.,   {van der
  Marel} R.~P.,  2013, \mn@doi [\apj] {10.1088/0004-637X/768/2/140}, \href
  {https://ui.adsabs.harvard.edu/abs/2013ApJ...768..140B} {768, 140}

\bibitem[\protect\citeauthoryear{{Busha}, {Wechsler}, {Behroozi}, {Gerke},
  {Klypin}  \& {Primack}}{{Busha} et~al.}{2011}]{Busha.etal.2011}
{Busha} M.~T.,  {Wechsler} R.~H.,  {Behroozi} P.~S.,  {Gerke} B.~F.,  {Klypin}
  A.~A.,   {Primack} J.~R.,  2011, \mn@doi [\apj]
  {10.1088/0004-637X/743/2/117}, \href
  {https://ui.adsabs.harvard.edu/abs/2011ApJ...743..117B} {743, 117}

\bibitem[\protect\citeauthoryear{{Callingham} et~al.,}{{Callingham}
  et~al.}{2019}]{Callingham.etal.2019}
{Callingham} T.~M.,  et~al., 2019, \mn@doi [\mnras] {10.1093/mnras/stz365},
  \href {https://ui.adsabs.harvard.edu/abs/2019MNRAS.484.5453C} {484, 5453}

\bibitem[\protect\citeauthoryear{{Conroy}, {Naidu}, {Garavito-Camargo},
  {Besla}, {Zaritsky}, {Bonaca}  \& {Johnson}}{{Conroy}
  et~al.}{2021}]{Conroy.etal.2021}
{Conroy} C.,  {Naidu} R.~P.,  {Garavito-Camargo} N.,  {Besla} G.,  {Zaritsky}
  D.,  {Bonaca} A.,   {Johnson} B.~D.,  2021, \mn@doi [\nat]
  {10.1038/s41586-021-03385-7}, \href
  {https://ui.adsabs.harvard.edu/abs/2021Natur.592..534C} {592, 534}

\bibitem[\protect\citeauthoryear{{Correa Magnus} \& {Vasiliev}}{{Correa Magnus}
  \& {Vasiliev}}{2022}]{Correa_Magnus.Vasiliev.2022}
{Correa Magnus} L.,  {Vasiliev} E.,  2022, \mn@doi [\mnras]
  {10.1093/mnras/stab3726}, \href
  {https://ui.adsabs.harvard.edu/abs/2022MNRAS.511.2610C} {511, 2610}

\bibitem[\protect\citeauthoryear{{D'Souza} \& {Bell}}{{D'Souza} \&
  {Bell}}{2022}]{DSouza.Bell.2022}
{D'Souza} R.,  {Bell} E.~F.,  2022, \mn@doi [\mnras] {10.1093/mnras/stac404},
  \href {https://ui.adsabs.harvard.edu/abs/2022MNRAS.512..739D} {512, 739}

\bibitem[\protect\citeauthoryear{{Deason} et~al.,}{{Deason}
  et~al.}{2021}]{Deason.etal.2021}
{Deason} A.~J.,  et~al., 2021, \mn@doi [\mnras] {10.1093/mnras/staa3984}, \href
  {https://ui.adsabs.harvard.edu/abs/2021MNRAS.501.5964D} {501, 5964}

\bibitem[\protect\citeauthoryear{{Diaferio}, {Geller}  \& {Rines}}{{Diaferio}
  et~al.}{2005}]{Diaferio.etal.2005}
{Diaferio} A.,  {Geller} M.~J.,   {Rines} K.~J.,  2005, \mn@doi [\apjl]
  {10.1086/432880}, \href
  {https://ui.adsabs.harvard.edu/abs/2005ApJ...628L..97D} {628, L97}

\bibitem[\protect\citeauthoryear{{Eadie} \& {Juri{\'c}}}{{Eadie} \&
  {Juri{\'c}}}{2019}]{Eadie.Juric.2019}
{Eadie} G.,  {Juri{\'c}} M.,  2019, \mn@doi [\apj] {10.3847/1538-4357/ab0f97},
  \href {https://ui.adsabs.harvard.edu/abs/2019ApJ...875..159E} {875, 159}

\bibitem[\protect\citeauthoryear{{Eadie}, {Springford}  \& {Harris}}{{Eadie}
  et~al.}{2017}]{Eadie.etal.2017}
{Eadie} G.~M.,  {Springford} A.,   {Harris} W.~E.,  2017, \mn@doi [\apj]
  {10.3847/1538-4357/835/2/167}, \href
  {https://ui.adsabs.harvard.edu/abs/2017ApJ...835..167E} {835, 167}

\bibitem[\protect\citeauthoryear{{Erkal} et~al.,}{{Erkal}
  et~al.}{2019}]{Erkal.etal.2019}
{Erkal} D.,  et~al., 2019, \mn@doi [\mnras] {10.1093/mnras/stz1371}, \href
  {https://ui.adsabs.harvard.edu/abs/2019MNRAS.487.2685E} {487, 2685}

\bibitem[\protect\citeauthoryear{{Erkal}, {Belokurov}  \& {Parkin}}{{Erkal}
  et~al.}{2020}]{Erkal.etal.2020}
{Erkal} D.,  {Belokurov} V.~A.,   {Parkin} D.~L.,  2020, \mn@doi [\mnras]
  {10.1093/mnras/staa2840}, \href
  {https://ui.adsabs.harvard.edu/abs/2020MNRAS.498.5574E} {498, 5574}

\bibitem[\protect\citeauthoryear{{Erkal} et~al.,}{{Erkal}
  et~al.}{2021}]{Erkal.etal.2021}
{Erkal} D.,  et~al., 2021, \mn@doi [\mnras] {10.1093/mnras/stab1828}, \href
  {https://ui.adsabs.harvard.edu/abs/2021MNRAS.506.2677E} {506, 2677}

\bibitem[\protect\citeauthoryear{{Feldmann}}{{Feldmann}}{2013}]{Feldmann.2013}
{Feldmann} R.,  2013, \mn@doi [\mnras] {10.1093/mnras/stt851}, \href
  {https://ui.adsabs.harvard.edu/abs/2013MNRAS.433.1910F} {433, 1910}

\bibitem[\protect\citeauthoryear{{Fritz}, {Di Cintio}, {Battaglia}, {Brook}  \&
  {Taibi}}{{Fritz} et~al.}{2020}]{Fritz.etal.2020}
{Fritz} T.~K.,  {Di Cintio} A.,  {Battaglia} G.,  {Brook} C.,   {Taibi} S.,
  2020, \mn@doi [\mnras] {10.1093/mnras/staa1040}, \href
  {https://ui.adsabs.harvard.edu/abs/2020MNRAS.494.5178F} {494, 5178}

\bibitem[\protect\citeauthoryear{{Garavito-Camargo}, {Besla}, {Laporte},
  {Johnston}, {G{\'o}mez}  \& {Watkins}}{{Garavito-Camargo}
  et~al.}{2019}]{Garavito.Camargo.etal.2019}
{Garavito-Camargo} N.,  {Besla} G.,  {Laporte} C. F.~P.,  {Johnston} K.~V.,
  {G{\'o}mez} F.~A.,   {Watkins} L.~L.,  2019, \mn@doi [\apj]
  {10.3847/1538-4357/ab32eb}, \href
  {https://ui.adsabs.harvard.edu/abs/2019ApJ...884...51G} {884, 51}

\bibitem[\protect\citeauthoryear{{Garavito-Camargo}, {Besla}, {Laporte},
  {Price-Whelan}, {Cunningham}, {Johnston}, {Weinberg}  \&
  {G{\'o}mez}}{{Garavito-Camargo} et~al.}{2021}]{Garavito-Camargo.etal.2021}
{Garavito-Camargo} N.,  {Besla} G.,  {Laporte} C. F.~P.,  {Price-Whelan} A.~M.,
   {Cunningham} E.~C.,  {Johnston} K.~V.,  {Weinberg} M.,   {G{\'o}mez} F.~A.,
  2021, \mn@doi [\apj] {10.3847/1538-4357/ac0b44}, \href
  {https://ui.adsabs.harvard.edu/abs/2021ApJ...919..109G} {919, 109}

\bibitem[\protect\citeauthoryear{{Garavito-Camargo} et~al.,}{{Garavito-Camargo}
  et~al.}{2023}]{Garavito.Camargo.etal.2023}
{Garavito-Camargo} N.,  et~al., 2023, \mn@doi [arXiv e-prints]
  {10.48550/arXiv.2311.11359}, \href
  {https://ui.adsabs.harvard.edu/abs/2023arXiv231111359G} {p. arXiv:2311.11359}

\bibitem[\protect\citeauthoryear{{Gibbons}, {Belokurov}  \& {Evans}}{{Gibbons}
  et~al.}{2014}]{Gibbons.etal.2014}
{Gibbons} S.~L.~J.,  {Belokurov} V.,   {Evans} N.~W.,  2014, \mn@doi [\mnras]
  {10.1093/mnras/stu1986}, \href
  {https://ui.adsabs.harvard.edu/abs/2014MNRAS.445.3788G} {445, 3788}

\bibitem[\protect\citeauthoryear{{G{\'o}mez}, {Besla}, {Carpintero},
  {Villalobos}, {O'Shea}  \& {Bell}}{{G{\'o}mez}
  et~al.}{2015}]{Gomez.etal.2015}
{G{\'o}mez} F.~A.,  {Besla} G.,  {Carpintero} D.~D.,  {Villalobos} {\'A}.,
  {O'Shea} B.~W.,   {Bell} E.~F.,  2015, \mn@doi [\apj]
  {10.1088/0004-637X/802/2/128}, \href
  {https://ui.adsabs.harvard.edu/abs/2015ApJ...802..128G} {802, 128}

\bibitem[\protect\citeauthoryear{{Gonz{\'a}lez}, {Kravtsov}  \&
  {Gnedin}}{{Gonz{\'a}lez} et~al.}{2013}]{Gonzalez.etal.2013}
{Gonz{\'a}lez} R.~E.,  {Kravtsov} A.~V.,   {Gnedin} N.~Y.,  2013, \mn@doi
  [\apj] {10.1088/0004-637X/770/2/96}, \href
  {https://ui.adsabs.harvard.edu/abs/2013ApJ...770...96G} {770, 96}

\bibitem[\protect\citeauthoryear{{Gonz{\'a}lez}, {Kravtsov}  \&
  {Gnedin}}{{Gonz{\'a}lez} et~al.}{2014}]{Gonzalez.etal.2014}
{Gonz{\'a}lez} R.~E.,  {Kravtsov} A.~V.,   {Gnedin} N.~Y.,  2014, \mn@doi
  [\apj] {10.1088/0004-637X/793/2/91}, \href
  {https://ui.adsabs.harvard.edu/abs/2014ApJ...793...91G} {793, 91}

\bibitem[\protect\citeauthoryear{{Griffen}, {Ji}, {Dooley}, {G{\'o}mez},
  {Vogelsberger}, {O'Shea}  \& {Frebel}}{{Griffen}
  et~al.}{2016}]{Griffen.etal.2016}
{Griffen} B.~F.,  {Ji} A.~P.,  {Dooley} G.~A.,  {G{\'o}mez} F.~A.,
  {Vogelsberger} M.,  {O'Shea} B.~W.,   {Frebel} A.,  2016, \mn@doi [\apj]
  {10.3847/0004-637X/818/1/10}, \href
  {https://ui.adsabs.harvard.edu/abs/2016ApJ...818...10G} {818, 10}

\bibitem[\protect\citeauthoryear{Hunter}{Hunter}{2007}]{matplotlib}
Hunter J.~D.,  2007, \mn@doi [Computing In Science \& Engineering]
  {10.1109/MCSE.2007.55}, 9, 90

\bibitem[\protect\citeauthoryear{{Ji} et~al.,}{{Ji}
  et~al.}{2021}]{Ji.etal.2021}
{Ji} A.~P.,  et~al., 2021, \mn@doi [\apj] {10.3847/1538-4357/ac1869}, \href
  {https://ui.adsabs.harvard.edu/abs/2021ApJ...921...32J} {921, 32}

\bibitem[\protect\citeauthoryear{Jones, Oliphant, Peterson  et~al.}{Jones
  et~al.}{01  }]{scipy}
Jones E.,  Oliphant T.,  Peterson P.,   et~al., 2001--, {SciPy}: Open source
  scientific tools for {Python}, \url {http://www.scipy.org/}

\bibitem[\protect\citeauthoryear{{Kahn} \& {Woltjer}}{{Kahn} \&
  {Woltjer}}{1959}]{Kahn.Woltjer.1959}
{Kahn} F.~D.,  {Woltjer} L.,  1959, \mn@doi [\apj] {10.1086/146762}, \href
  {https://ui.adsabs.harvard.edu/abs/1959ApJ...130..705K} {130, 705}

\bibitem[\protect\citeauthoryear{{Karachentsev} \& {Kashibadze}}{{Karachentsev}
  \& {Kashibadze}}{2006}]{Karachentsev.Kashibadze.2006}
{Karachentsev} I.~D.,  {Kashibadze} O.~G.,  2006, \mn@doi [Astrophysics]
  {10.1007/s10511-006-0002-6}, \href
  {https://ui.adsabs.harvard.edu/abs/2006Ap.....49....3K} {49, 3}

\bibitem[\protect\citeauthoryear{{Kochanek}}{{Kochanek}}{1996}]{Kochanek.1996}
{Kochanek} C.~S.,  1996, \mn@doi [\apj] {10.1086/176724}, \href
  {https://ui.adsabs.harvard.edu/abs/1996ApJ...457..228K} {457, 228}

\bibitem[\protect\citeauthoryear{{Koposov} et~al.,}{{Koposov}
  et~al.}{2023}]{Koposov.etal.2023}
{Koposov} S.~E.,  et~al., 2023, \mn@doi [\mnras] {10.1093/mnras/stad551}, \href
  {https://ui.adsabs.harvard.edu/abs/2023MNRAS.521.4936K} {521, 4936}

\bibitem[\protect\citeauthoryear{{Kravtsov} \& {Manwadkar}}{{Kravtsov} \&
  {Manwadkar}}{2022}]{Kravtsov.Manwadkar.2022}
{Kravtsov} A.,  {Manwadkar} V.,  2022, \mn@doi [\mnras]
  {10.1093/mnras/stac1439}, \href
  {https://ui.adsabs.harvard.edu/abs/2022MNRAS.514.2667K} {514, 2667}

\bibitem[\protect\citeauthoryear{{Krumholz} \& {Dekel}}{{Krumholz} \&
  {Dekel}}{2012}]{Krumholz.Dekel.2012}
{Krumholz} M.~R.,  {Dekel} A.,  2012, \mn@doi [\apj]
  {10.1088/0004-637X/753/1/16}, \href
  {https://ui.adsabs.harvard.edu/abs/2012ApJ...753...16K} {753, 16}

\bibitem[\protect\citeauthoryear{{Li} \& {White}}{{Li} \&
  {White}}{2008}]{Li.White.2008}
{Li} Y.-S.,  {White} S. D.~M.,  2008, \mn@doi [\mnras]
  {10.1111/j.1365-2966.2007.12748.x}, \href
  {https://ui.adsabs.harvard.edu/abs/2008MNRAS.384.1459L} {384, 1459}

\bibitem[\protect\citeauthoryear{{Li}, {Qian}, {Han}, {Li}, {Wang}  \&
  {Jing}}{{Li} et~al.}{2020}]{Li.etal.2020}
{Li} Z.-Z.,  {Qian} Y.-Z.,  {Han} J.,  {Li} T.~S.,  {Wang} W.,   {Jing} Y.~P.,
  2020, \mn@doi [\apj] {10.3847/1538-4357/ab84f0}, \href
  {https://ui.adsabs.harvard.edu/abs/2020ApJ...894...10L} {894, 10}

\bibitem[\protect\citeauthoryear{{Li}, {Hammer}, {Babusiaux}, {Pawlowski},
  {Yang}, {Arenou}, {Du}  \& {Wang}}{{Li} et~al.}{2021}]{Li.etal.2021}
{Li} H.,  {Hammer} F.,  {Babusiaux} C.,  {Pawlowski} M.~S.,  {Yang} Y.,
  {Arenou} F.,  {Du} C.,   {Wang} J.,  2021, \mn@doi [\apj]
  {10.3847/1538-4357/ac0436}, \href
  {https://ui.adsabs.harvard.edu/abs/2021ApJ...916....8L} {916, 8}

\bibitem[\protect\citeauthoryear{{Lilly}, {Carollo}, {Pipino}, {Renzini}  \&
  {Peng}}{{Lilly} et~al.}{2013}]{Lilly.etal.2013}
{Lilly} S.~J.,  {Carollo} C.~M.,  {Pipino} A.,  {Renzini} A.,   {Peng} Y.,
  2013, \mn@doi [\apj] {10.1088/0004-637X/772/2/119}, \href
  {https://ui.adsabs.harvard.edu/abs/2013ApJ...772..119L} {772, 119}

\bibitem[\protect\citeauthoryear{{Makarov}, {Khoperskov}, {Makarov},
  {Makarova}, {Libeskind}  \& {Salomon}}{{Makarov}
  et~al.}{2023}]{Makarov.etal.2023}
{Makarov} D.,  {Khoperskov} S.,  {Makarov} D.,  {Makarova} L.,  {Libeskind} N.,
    {Salomon} J.-B.,  2023, \mn@doi [\mnras] {10.1093/mnras/stad757}, \href
  {https://ui.adsabs.harvard.edu/abs/2023MNRAS.521.3540M} {521, 3540}

\bibitem[\protect\citeauthoryear{{Manwadkar} \& {Kravtsov}}{{Manwadkar} \&
  {Kravtsov}}{2022}]{Manwadkar.Kravtsov.2022}
{Manwadkar} V.,  {Kravtsov} A.~V.,  2022, \mn@doi [\mnras]
  {10.1093/mnras/stac2452}, \href
  {https://ui.adsabs.harvard.edu/abs/2022MNRAS.516.3944M} {516, 3944}

\bibitem[\protect\citeauthoryear{{McConnachie} \& {Venn}}{{McConnachie} \&
  {Venn}}{2020a}]{McConnachie.Venn.2020b}
{McConnachie} A.~W.,  {Venn} K.~A.,  2020a, \mn@doi [Research Notes of the
  American Astronomical Society] {10.3847/2515-5172/abd18b}, \href
  {https://ui.adsabs.harvard.edu/abs/2020RNAAS...4..229M} {4, 229}

\bibitem[\protect\citeauthoryear{McConnachie \& Venn}{McConnachie \&
  Venn}{2020b}]{McConnachie.2020a}
McConnachie A.~W.,  Venn K.~A.,  2020b, \mn@doi [Research Notes of the AAS]
  {10.3847/2515-5172/abd18b}, 4, 229

\bibitem[\protect\citeauthoryear{{McConnachie} \& {Venn}}{{McConnachie} \&
  {Venn}}{2020c}]{McConnachie.Venn.2020}
{McConnachie} A.~W.,  {Venn} K.~A.,  2020c, \mn@doi [\aj]
  {10.3847/1538-3881/aba4ab}, \href
  {https://ui.adsabs.harvard.edu/abs/2020AJ....160..124M} {160, 124}

\bibitem[\protect\citeauthoryear{McConnachie \& Venn}{McConnachie \&
  Venn}{2020d}]{McConnachie.2020b}
McConnachie A.~W.,  Venn K.~A.,  2020d, \mn@doi [The Astronomical Journal]
  {10.3847/1538-3881/aba4ab}, 160, 124

\bibitem[\protect\citeauthoryear{{Nadler} et~al.,}{{Nadler}
  et~al.}{2020}]{Nadler.etal.2020}
{Nadler} E.~O.,  et~al., 2020, \mn@doi [\apj] {10.3847/1538-4357/ab846a}, \href
  {https://ui.adsabs.harvard.edu/abs/2020ApJ...893...48N} {893, 48}

\bibitem[\protect\citeauthoryear{{Pace}, {Erkal}  \& {Li}}{{Pace}
  et~al.}{2022}]{Pace.etal.2022}
{Pace} A.~B.,  {Erkal} D.,   {Li} T.~S.,  2022, \mn@doi [\apj]
  {10.3847/1538-4357/ac997b}, \href
  {https://ui.adsabs.harvard.edu/abs/2022ApJ...940..136P} {940, 136}

\bibitem[\protect\citeauthoryear{{Patel}, {Besla}  \& {Sohn}}{{Patel}
  et~al.}{2017a}]{Patel.etal.2017a}
{Patel} E.,  {Besla} G.,   {Sohn} S.~T.,  2017a, \mn@doi [\mnras]
  {10.1093/mnras/stw2616}, \href
  {https://ui.adsabs.harvard.edu/abs/2017MNRAS.464.3825P} {464, 3825}

\bibitem[\protect\citeauthoryear{{Patel}, {Besla}  \& {Mandel}}{{Patel}
  et~al.}{2017b}]{Patel.etal.2017b}
{Patel} E.,  {Besla} G.,   {Mandel} K.,  2017b, \mn@doi [\mnras]
  {10.1093/mnras/stx698}, \href
  {https://ui.adsabs.harvard.edu/abs/2017MNRAS.468.3428P} {468, 3428}

\bibitem[\protect\citeauthoryear{{Patel}, {Besla}, {Mandel}  \& {Sohn}}{{Patel}
  et~al.}{2018}]{Patel.etal.2018}
{Patel} E.,  {Besla} G.,  {Mandel} K.,   {Sohn} S.~T.,  2018, \mn@doi [\apj]
  {10.3847/1538-4357/aab78f}, \href
  {https://ui.adsabs.harvard.edu/abs/2018ApJ...857...78P} {857, 78}

\bibitem[\protect\citeauthoryear{{Pe{\~n}arrubia}, {G{\'o}mez}, {Besla},
  {Erkal}  \& {Ma}}{{Pe{\~n}arrubia} et~al.}{2016}]{Penarrubia.etal.2016}
{Pe{\~n}arrubia} J.,  {G{\'o}mez} F.~A.,  {Besla} G.,  {Erkal} D.,   {Ma}
  Y.-Z.,  2016, \mn@doi [\mnras] {10.1093/mnrasl/slv160}, \href
  {https://ui.adsabs.harvard.edu/abs/2016MNRAS.456L..54P} {456, L54}

\bibitem[\protect\citeauthoryear{{Petersen} \& {Pe{\~n}arrubia}}{{Petersen} \&
  {Pe{\~n}arrubia}}{2020}]{Petersen.Penarrubia.2020}
{Petersen} M.~S.,  {Pe{\~n}arrubia} J.,  2020, \mn@doi [\mnras]
  {10.1093/mnrasl/slaa029}, \href
  {https://ui.adsabs.harvard.edu/abs/2020MNRAS.494L..11P} {494, L11}

\bibitem[\protect\citeauthoryear{{Petersen} \& {Pe{\~n}arrubia}}{{Petersen} \&
  {Pe{\~n}arrubia}}{2021}]{Petersen.Penarrubia.2021}
{Petersen} M.~S.,  {Pe{\~n}arrubia} J.,  2021, \mn@doi [Nature Astronomy]
  {10.1038/s41550-020-01254-3}, \href
  {https://ui.adsabs.harvard.edu/abs/2021NatAs...5..251P} {5, 251}

\bibitem[\protect\citeauthoryear{{Pham}, {Kravtsov}  \& {Manwadkar}}{{Pham}
  et~al.}{2023}]{Pham.etal.2023}
{Pham} K.,  {Kravtsov} A.,   {Manwadkar} V.,  2023, \mn@doi [\mnras]
  {10.1093/mnras/stad335}, \href
  {https://ui.adsabs.harvard.edu/abs/2023MNRAS.520.3937P} {520, 3937}

\bibitem[\protect\citeauthoryear{{Posti} \& {Helmi}}{{Posti} \&
  {Helmi}}{2019}]{Posti.Helmi.2019}
{Posti} L.,  {Helmi} A.,  2019, \mn@doi [\aap] {10.1051/0004-6361/201833355},
  \href {https://ui.adsabs.harvard.edu/abs/2019A&A...621A..56P} {621, A56}

\bibitem[\protect\citeauthoryear{{Roche}, {Necib}, {Lin}, {Ou}  \&
  {Nguyen}}{{Roche} et~al.}{2024}]{Roche.etal.2024}
{Roche} C.,  {Necib} L.,  {Lin} T.,  {Ou} X.,   {Nguyen} T.,  2024, \mn@doi
  [arXiv e-prints] {10.48550/arXiv.2402.00108}, \href
  {https://ui.adsabs.harvard.edu/abs/2024arXiv240200108R} {p. arXiv:2402.00108}

\bibitem[\protect\citeauthoryear{{Rodriguez Wimberly} et~al.,}{{Rodriguez
  Wimberly} et~al.}{2022}]{Rodriguez_Wimberly_etal2022}
{Rodriguez Wimberly} M.~K.,  et~al., 2022, \mn@doi [\mnras]
  {10.1093/mnras/stac1265}, \href
  {https://ui.adsabs.harvard.edu/abs/2022MNRAS.513.4968R} {513, 4968}

\bibitem[\protect\citeauthoryear{{Sandage}}{{Sandage}}{1986}]{Sandage.1986}
{Sandage} A.,  1986, \mn@doi [\apj] {10.1086/164387}, \href
  {https://ui.adsabs.harvard.edu/abs/1986ApJ...307....1S} {307, 1}

\bibitem[\protect\citeauthoryear{{Sheng}, {Ting}, {Xue}, {Chang}  \&
  {Tian}}{{Sheng} et~al.}{2024}]{Sheng.etal.2024}
{Sheng} Y.,  {Ting} Y.-S.,  {Xue} X.-X.,  {Chang} J.,   {Tian} H.,  2024,
  \mn@doi [arXiv e-prints] {10.48550/arXiv.2404.08975}, \href
  {https://ui.adsabs.harvard.edu/abs/2024arXiv240408975S} {p. arXiv:2404.08975}

\bibitem[\protect\citeauthoryear{{Shipp} et~al.,}{{Shipp}
  et~al.}{2021}]{Shipp.etal.2021}
{Shipp} N.,  et~al., 2021, \mn@doi [\apj] {10.3847/1538-4357/ac2e93}, \href
  {https://ui.adsabs.harvard.edu/abs/2021ApJ...923..149S} {923, 149}

\bibitem[\protect\citeauthoryear{{Simon}}{{Simon}}{2019}]{Simon.2019}
{Simon} J.~D.,  2019, \mn@doi [\araa] {10.1146/annurev-astro-091918-104453},
  \href {https://ui.adsabs.harvard.edu/abs/2019ARA&A..57..375S} {57, 375}

\bibitem[\protect\citeauthoryear{{Smith}}{{Smith}}{1936}]{Smith.1936}
{Smith} S.,  1936, \mn@doi [\apj] {10.1086/143697}, \href
  {https://ui.adsabs.harvard.edu/abs/1936ApJ....83...23S} {83, 23}

\bibitem[\protect\citeauthoryear{{Sohn}, {Besla}, {van der Marel},
  {Boylan-Kolchin}, {Majewski}  \& {Bullock}}{{Sohn}
  et~al.}{2013}]{Sohn.etal.2013}
{Sohn} S.~T.,  {Besla} G.,  {van der Marel} R.~P.,  {Boylan-Kolchin} M.,
  {Majewski} S.~R.,   {Bullock} J.~S.,  2013, \mn@doi [\apj]
  {10.1088/0004-637X/768/2/139}, \href
  {https://ui.adsabs.harvard.edu/abs/2013ApJ...768..139S} {768, 139}

\bibitem[\protect\citeauthoryear{{Torrealba} et~al.,}{{Torrealba}
  et~al.}{2019}]{Torrealba.2019}
{Torrealba} G.,  et~al., 2019, \mn@doi [\mnras] {10.1093/mnras/stz1624}, \href
  {https://ui.adsabs.harvard.edu/abs/2019MNRAS.488.2743T} {488, 2743}

\bibitem[\protect\citeauthoryear{{Van Der Walt}, {Colbert}  \&
  {Varoquaux}}{{Van Der Walt} et~al.}{2011}]{numpy_ndarray}
{Van Der Walt} S.,  {Colbert} S.~C.,   {Varoquaux} G.,  2011, ArXiv:1102.1523,
  \href {http://adsabs.harvard.edu/abs/2011arXiv1102.1523V} {p.~1}

\bibitem[\protect\citeauthoryear{{Vasiliev}}{{Vasiliev}}{2019}]{Vasiliev.2019}
{Vasiliev} E.,  2019, \mn@doi [\mnras] {10.1093/mnras/stz171}, \href
  {https://ui.adsabs.harvard.edu/abs/2019MNRAS.484.2832V} {484, 2832}

\bibitem[\protect\citeauthoryear{{Vasiliev}}{{Vasiliev}}{2023}]{Vasiliev.2023}
{Vasiliev} E.,  2023, \mn@doi [Galaxies] {10.3390/galaxies11020059}, \href
  {https://ui.adsabs.harvard.edu/abs/2023Galax..11...59V} {11, 59}

\bibitem[\protect\citeauthoryear{{Vasiliev}, {Belokurov}  \&
  {Erkal}}{{Vasiliev} et~al.}{2021}]{Vasiliev.etal.2021}
{Vasiliev} E.,  {Belokurov} V.,   {Erkal} D.,  2021, \mn@doi [\mnras]
  {10.1093/mnras/staa3673}, \href
  {https://ui.adsabs.harvard.edu/abs/2021MNRAS.501.2279V} {501, 2279}

\bibitem[\protect\citeauthoryear{{Watkins}, {Evans}  \& {An}}{{Watkins}
  et~al.}{2010}]{Watkins.etal.2010}
{Watkins} L.~L.,  {Evans} N.~W.,   {An} J.~H.,  2010, \mn@doi [\mnras]
  {10.1111/j.1365-2966.2010.16708.x}, \href
  {https://ui.adsabs.harvard.edu/abs/2010MNRAS.406..264W} {406, 264}

\bibitem[\protect\citeauthoryear{{Watkins}, {van der Marel}  \&
  {Bennet}}{{Watkins} et~al.}{2024}]{Watkins.etal.2024}
{Watkins} L.~L.,  {van der Marel} R.~P.,   {Bennet} P.,  2024, \mn@doi [\apj]
  {10.3847/1538-4357/ad1f58}, \href
  {https://ui.adsabs.harvard.edu/abs/2024ApJ...963...84W} {963, 84}

\bibitem[\protect\citeauthoryear{Weisz et~al.,}{Weisz
  et~al.}{2016}]{Weisz.2016}
Weisz D.~R.,  et~al., 2016, \mn@doi [The Astrophysical Journal]
  {10.3847/0004-637X/822/1/32}, 822, 32

\bibitem[\protect\citeauthoryear{{Yaaqib}, {Petersen}  \&
  {Pe{\~n}arrubia}}{{Yaaqib} et~al.}{2024}]{Yaaqib.etal.2024}
{Yaaqib} R.,  {Petersen} M.~S.,   {Pe{\~n}arrubia} J.,  2024, \mn@doi [\mnras]
  {10.1093/mnras/stae1363}, \href
  {https://ui.adsabs.harvard.edu/abs/2024MNRAS.tmp.1351Y} {}

\bibitem[\protect\citeauthoryear{{Zaritsky}, {Olszewski}, {Schommer},
  {Peterson}  \& {Aaronson}}{{Zaritsky} et~al.}{1989}]{Zaritsky.etal.1989}
{Zaritsky} D.,  {Olszewski} E.~W.,  {Schommer} R.~A.,  {Peterson} R.~C.,
  {Aaronson} M.,  1989, \mn@doi [\apj] {10.1086/167947}, \href
  {https://ui.adsabs.harvard.edu/abs/1989ApJ...345..759Z} {345, 759}

\bibitem[\protect\citeauthoryear{{Zhou}, {Li}, {Huang}  \& {Zhang}}{{Zhou}
  et~al.}{2023}]{Zhou.etal.2023}
{Zhou} Y.,  {Li} X.,  {Huang} Y.,   {Zhang} H.,  2023, \mn@doi [\apj]
  {10.3847/1538-4357/acadd9}, \href
  {https://ui.adsabs.harvard.edu/abs/2023ApJ...946...73Z} {946, 73}

\bibitem[\protect\citeauthoryear{{Zwicky}}{{Zwicky}}{1933}]{Zwicky.1933}
{Zwicky} F.,  1933, Helvetica Physica Acta, \href
  {https://ui.adsabs.harvard.edu/abs/1933AcHPh...6..110Z} {6, 110}

\bibitem[\protect\citeauthoryear{{Zwicky}}{{Zwicky}}{1937}]{Zwicky.1937}
{Zwicky} F.,  1937, \mn@doi [\apj] {10.1086/143864}, \href
  {https://ui.adsabs.harvard.edu/abs/1937ApJ....86..217Z} {86, 217}

\makeatother
\end{thebibliography}

\end{document}